\documentclass[%
 reprint,
 amsmath,amssymb,
 aps, superscriptaddress,
 longbibliography
]{revtex4-1}

\usepackage{eurosym}
\usepackage{amsmath}
\usepackage{graphicx}
\usepackage{dcolumn}
\usepackage{bm}
\usepackage{mathtools}
\usepackage{xcolor}
\usepackage{ulem}
\usepackage{setspace}
\usepackage{svg}

\begin{document}

\preprint{APS/123-QED}

\title{Gate control of magnon spin transport in unconventional magnon transistors based on the van der Waals antiferromagnet CrPS\textsubscript{4}}
\author{Dennis K. de Wal}\email{d.k.de.wal@rug.nl}
\author{Ra\'ul Luna Mena}
\author{Muhammad Zohaib}
\author {Bart J. van Wees}\address{Zernike Institute for Advanced Materials, University of Groningen, Nijenborgh 3, 9747 AG, Groningen, the Netherlands}
\date{November 21, 2024 }

\begin{abstract}
Magnon based spintronic devices require the modulation of magnon spin transport for their operations. We provide a proof-of-principle of an unconventional two-terminal non-local magnon transport device where the same contact is simultaneously used as injector and gate and an unconventional three-terminal non-local magnon transport device where the gate is positioned outside the magnon transport channel. In these devices we modulate the diffusive magnon transport of incoherent magnons in the van der Waals antiferromagnet Chromium thiophosphate, CrPS\textsubscript{4}. The non-local signals generated electrically by spin injection via the spin Hall effect (SHE) and thermally via the spin Seebeck effect (SSE) are altered by a modulator electrode. The current through the modulator increases or decreases the magnon chemical potential via the SHE and changes the magnon temperature through Joule heating. We achieve up to $\eta^{1\omega}_{\scriptscriptstyle{SHE}}=25$ $\%/$mA and $\eta^{2\omega}_{\scriptscriptstyle{SHE}}=16$ $\%/$mA modulation efficiencies for the electrically and thermally generated magnon spin transport, respectively, for CrPS\textsubscript{4} in the collinear state at in plane fields $>$7 T at a temperature of 25 K. 
\end{abstract}

\maketitle


\section{Introduction}
In information processing technology, encoding, transport, and manipulation is crucial for its operation. Spin wave (magnon) based computing has been shown as a very suitable alternative to CMOS based electronics for encoding\cite{kruglyak_magnonics_2010,serga_YIG_2010,khitun_magnonic_2010} and transport in three-dimensional ferro, ferri and antiferromagnets\cite{das_anisotropic_2022, lebrun_long-distance_2020,wimmer_observation_2020,cornelissen_long-distance_2015}. Also on investigating the novel two-dimensional (2D) van der Waals magnetic materials, significant efforts have been made with regards to thermally generated\cite{xing_magnon_2019, feringa_spin-flop_2022,qi_giant_2023} and electrically generated magnon transport\cite{de_wal_long-distance_2023}. Yet, electrical control over and manipulation of magnon signals in these systems remains a challenge. In Yttrium Iron Garnet (YIG), the ``workhorse'' of magnonics, a magnon transistor in a conventional three-terminal non-local geometry is demonstrated and explored, showing that the magnon spin conductivity can be modulated both electrically and thermally\cite{cornelissen_spin-current-controlled_2018,guckelhorn_quantitative_2020,liu_electrically_2021}. In this geometry strong signal modulation is achieved up to 40 $\%/$mA\cite{liu_electrically_2021}.

Efficient and scalable control over magnon spin transport in 2D van der Waals materials is also crucial for achieving a controllable 2D magnon gas. In the van der Waals magnets advances have been made in magnon valves, based on the non-local geometry, in MnPS\textsubscript{3}\cite{chen_electrically_2021} and CrPS\textsubscript{4}\cite{qi_giant_2023}. Nevertheless, for both these experiments the magnon currents are generated thermally, by the spin Seebeck effect (SSE)\cite{uchida_observation_2008}, for which only convoluted information about the magnon transport properties, such as magnon relaxation length and magnon conductivity, can be obtained. Also, in these works, neither via the SHE injected magnon transport, nor modulation of the magnon transport via the SHE, are reported. Moreover, the reported 'off' state, explained as the zero-crossing of the non-local spin Seebeck effect (nl-SSE) voltages at specific gate DC-currents is, in fact, the sign change in magnon chemical potential at the detector as a function of thermal gradient, which is observed in YIG as a function of temperature, Joule heating and injector-detector spacing\cite{shan_influence_2016,cornelissen_nonlocal_2017}. Although the proof of principle for these thermally controllable magnon transistors in \cite{chen_electrically_2021} and \cite{qi_giant_2023} is highly relevant, only thermal control over the thermally generated magnon spin current, driven via Joule heating, is shown and not via the spin Hall effect (SHE)\cite{hirsch_spin_1999}. 

``All-electrical'' non-local magnon transport, where the magnon spin is injected via the SHE, has been shown in the 2D van der Waals antiferromagnet CrPS\textsubscript{4}\cite{de_wal_long-distance_2023, wal_magnon_2024}, making this material very suitable for fully electrical (via the SHE\cite{saitoh_conversion_2006, kajiwara_transmission_2010, zhang_spin_2012, zhang_magnon_2012}) modulation of magnon spin transport. CrPS\textsubscript{4} is an A-type antiferromagnet with out-of-plane anisotropy and has a Neél temperature of T$_{N}=38$ K\cite{peng_magnetic_2020}. In this work, we demonstrate and explain the working of a magnon transistor based on CrPS\textsubscript{4}, similar to the magnon transistor on YIG\cite{cornelissen_spin-current-controlled_2018}.  We elaborate on the effects on the magnon conductivity of electrical gating via the SHE and thermal gating via both injection of heat and through the SSE. which affect both the electrically injected and thermally injected magnon currents. Furthermore, we propose a two-terminal and an unconventional three-terminal non-local geometry instead of using the conventional three-terminal with the gate in the middle, to achieve a greater tunability of the magnon current.

\begin{figure}[b]
\includegraphics[width=\linewidth]{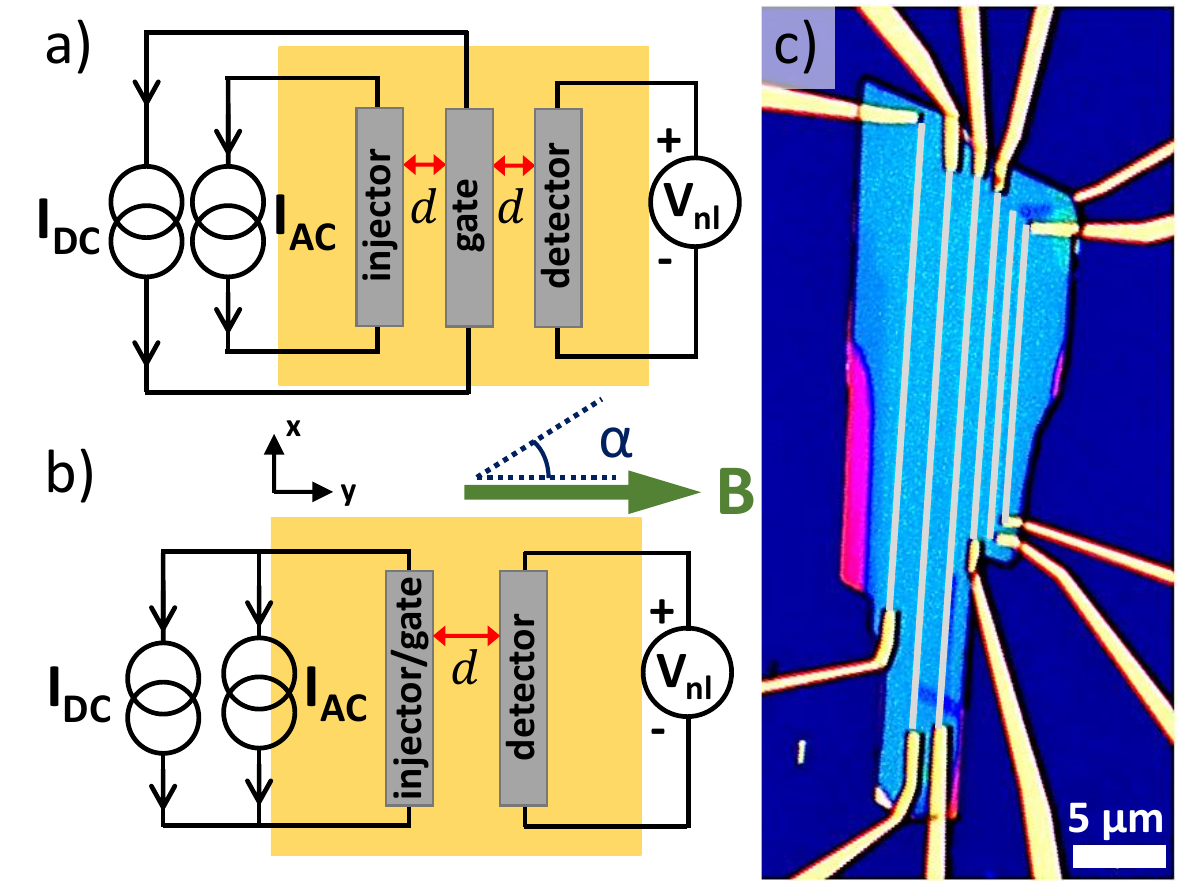}
\caption{Device D1: Schematic of the conventional three terminal (a) and the unconventional two terminal magnon transistor (b). Optical micrograph of device D1 (c). The green arrow represents the in-plane magnetic field and $\alpha$ is the angle of the in-plane rotation of the field. The picture of the device shows the Pt strips on top of a $\sim$100 nm thick CrPS\textsubscript{4} flake, which are contacted by Ti/Au lead contacts.}
\label{fig1:Devicegeometry}
\end{figure}

\section{Experimental concepts}
For electron transport in metals and semiconductors, the electron conductivity ($\sigma_e$) depends on the free electron density ($n_e$). This Drude model for electrons follows: $\sigma_e=e^2n_e\tau_e/m_e$, where $e$, $m_e$, and $\tau_e$ are the electron charge, effective mass, and the scattering time, respectively\cite{drude_zur_1900, drude_zur_1900-1}. For magnons in a system at finite temperature (thermal equilibrium magnons) we can define a similar relation. For out-of-equilibrium magnons, such as electrically injected and thermally generated (SSE) magnons, the magnon spin conductivity becomes:
\begin{equation}
    \sigma_m=\hbar\frac{n_m\tau_m}{m_m},
    \label{eq:Drude_magnon}
\end{equation}
where $n_m$ is the magnon density, which depends on both the magnon chemical potential as well as the temperature, $\tau_m$ is the magnon scattering time and $m_m=\hbar^2/(2J_S)$ is the effective mass with $J_S$ as the spin wave stiffness\cite{cornelissen_magnon_2016}. Therefore, $\sigma_m$ can be directly tuned via $n_m$.


\begin{figure*}[ptb]
\centering
\includegraphics[width=0.99\textwidth]{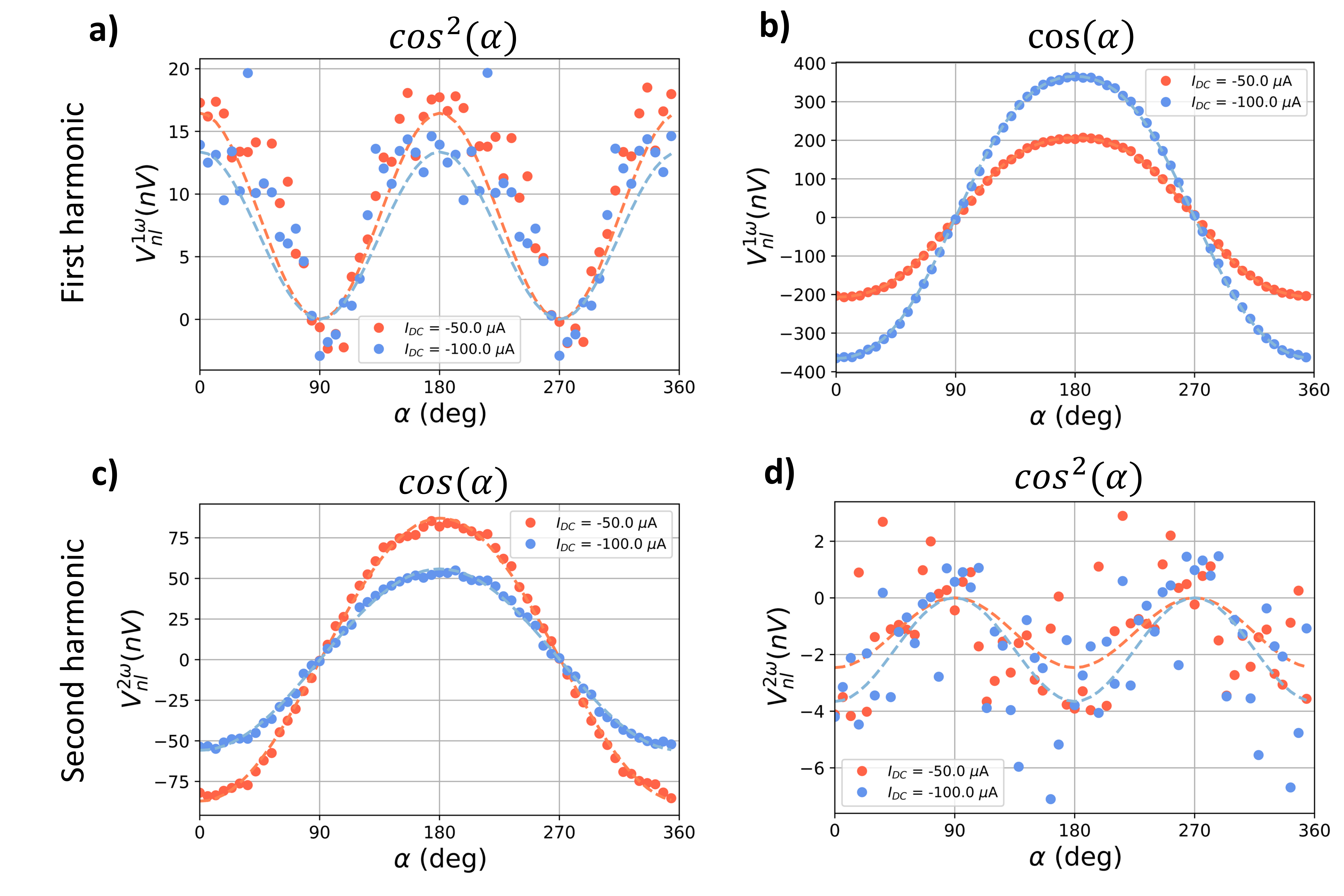}
\caption{First and second harmonic non-local voltages for $I_{AC}=60$ $\mu$A and $I_{DC}=-50$ $\mu$A and $I_{DC}=-100$ $\mu$A, as a function of $\alpha$ at a field of 7.5 T, T = 25 K. The symmetrized (around $\alpha=90^{\circ}$, see SI\cite{Supplementary_Mat_2024}) [antisymmetrized] data for the first harmonic is presented in figure (a) [(b)], the dashed lines are $\cos^2{\alpha}$ [$\cos{\alpha}$] fits to the data. The antisymmetrized [symmetrized] data for the second harmonic is presented in figure (c) [d], the dashed lines are $\cos{\alpha}$ [$\cos^2{\alpha}$] fits to the data. In figure (a) a constant offset of $V^{1\omega}_{nl}=2$nV is subtracted. For figure (b) the $D\cos{\alpha}$ cross term is much larger than the $B^{1\omega}\cos^3{\alpha}$ term.}
\label{fig2:AC_DC}%
\end{figure*}

In transistors based on magnons, spin is carried by magnons. A current through a gate contact affects the magnon transport in three ways: 1. The current generates an electronic spin accumulation ($\boldsymbol{\mu}_{mod}$) at the Pt/CrPS\textsubscript{4} interface via the SHE. By transfer of spin from the gate contact to the magnetic insulator, the magnon chemical potential $\mu_m$ is enhanced or depleted, in case the magnetization of CrPS\textsubscript{4} is collinear to $\boldsymbol{\mu}_{mod}$. Since a change in $\mu_{m}$ changes $n_m$, hence it also changes $\sigma_m$ between the injector and detector contacts. 2. The gate current generates heat by Joule heating, this alters the magnon temperature in the area of the CrPS\textsubscript{4} flake in proximity to the gate, changing $n_m$ via $T$. 3. Moreover, the increased temperature creates thermal gradients in the sample as well. These thermal gradients drive magnon currents via the SSE and therefore can lead to a change in $\mu_m$.  

We studied two Pt/CrPS\textsubscript{4} heterostructures (Device D1 and D2) where the CrPS\textsubscript{4} flakes are exfoliated from bulk crystals of CrPS\textsubscript{4} (HQgraphene). The 7 nm thick Pt electrodes are sputtered on top of a $\sim$100 nm thick flake. The contact spacing for the devices varies between 270 nm to 1400 nm edge-to-edge distance (see Fig. \ref{fig1:Devicegeometry}c) and the contacts have equal width ($\sim300$ nm). The length of the Pt strips is 20-40 $\mu$m. The Pt strips are contacted with Ti/Au leads to make electrical connections to the device. Angular dependent magnetoresistance (ADMR) measurements are performed for this non-local geometry as a function of in-plane angle $\alpha$ of the applied magnetic field w.r.t. the Pt strips (see Fig. \ref{fig1:Devicegeometry}). A low frequency ($\omega/(2\pi)<20$ Hz) ac-current $I=I_0\sin{\omega t}$ is applied to the injector Pt strip. The first ($V^{1\omega}$) and second ($V^{2\omega}$) harmonic voltage responses are measured at the detector Pt strip. All CrPS\textsubscript{4} flakes are exfoliated from the same bulk crystal.

First considering the conventional three-terminal magnon transistor, as given in Fig. \ref{fig1:Devicegeometry}a, we explore the effects of the modulation on the non-local magnon spin transport. The non-local voltages $V^{1\omega}_{nl}$ and $V^{2\omega}_{nl}$ in the detector correspond to the electrically and thermally generated magnon transport excited by the AC-current ($I_{AC}$) in the injector. Additionally, a modulating DC-current ($I_{DC}$) is applied to the gate electrode, affecting the magnon spin transport.
$V^{1\omega}$ and $V^{2\omega}$ will not be offsetted by $I_{DC}$ as we employ the lock-in method. We can summarize the effect on both $V^{1\omega}$ and $V^{2\omega}$ by:
\begin{align}
    V^{1\omega}     &=C_1I_{AC}\sigma_{m}(\alpha)\cos^2{(\alpha)},
    \label{eq:V1modulated}\\
    V^{2\omega}     &=C_2I^2_{AC}\sigma_{m}(\alpha)\cos{(\alpha)},
    \label{eq:V2modulated}
\end{align}
where $C_1$ and $C_2$ are the constants capturing the conversion of the charge currents to spin currents in the injector and detector\cite{cornelissen_spin-current-controlled_2018,de_wal_long-distance_2023}, for the electrical and thermal injection, respectively. The magnon conductivity $\sigma_m$ depends on $I_{DC}$ and is given by:
\begin{equation}
    \sigma_m(\alpha) = \sigma_m^0 + \Delta\sigma_JI^2_{DC} + \Delta\sigma_{SHE}I_{DC}\cos{(\alpha)}.
    \label{eq:sigma_m}
\end{equation}
Here $\sigma_m^0$ is the spin conductivity without any modulation by $I_{DC}$, $\Delta\sigma_J$ is the efficiency of modulation by Joule heating and $\Delta\sigma_{SHE}$ for the magnons injected by the SHE. For the latter, the injection depends on the col-linearity of $\boldsymbol{\mu}_{mod}$ and the net magnetization of the CrPS\textsubscript{4} ($\mathbf{m}=(\mathbf{m_1}+\mathbf{m_2})/2$, $\mathbf{m_{1,2}}$ are the sub-lattice magnetizations) via the SHE, as is the case for equation (\ref{eq:V1modulated}) and \ref{eq:V2modulated}.

Substituting $\sigma_m$ in equation (\ref{eq:V1modulated}) and (\ref{eq:V2modulated}) we arrive at the following responses:
\begin{align}
    V^{1\omega}     &=A^{1\omega}\cos^2{(\alpha)} + B^{1\omega}\cos^3{(\alpha)},
    \label{eq:V1angular}\\
    V^{2\omega}     &=A^{2\omega}\cos{(\alpha)} + B^{2\omega}\cos^2{(\alpha)},
    \label{eq:V2angular}
\end{align}
for which the Joule heating affects the amplitudes $A^{1\omega(2\omega)}$, scaling with $I_{DC}^2$, and the injected magnons via the SHE modify the amplitude $B^{1\omega(2\omega)}$, scaling with $I_{DC}$. 


\section{Results}
For our experiment employing the conventional three terminal magnon transistor (as given in Fig. \ref{fig1:Devicegeometry}a) we could not observe a signal for $V^{1\omega}_{nl}$ above the noise level of typically 5 nV\textsubscript{RMS}. This is caused by two factors: Firstly, the injector-detector distance is comparable to the magnon relaxation length $\lambda_m$, meaning the magnons can already decay before they reach the detector. Secondly, the Pt gate contact in between the injector and detector contact functions as a spin sink, absorbing magnon spin by spin flip scattering in the Pt at the Pt/CrPS\textsubscript{4} interface. Therefore, we combined the injector and gate into one contact, making a two terminal transistor (see Fig. \ref{fig1:Devicegeometry}b), with injector-detector spacing $d=340$ nm. As the gate affects the magnon density over a distance $\lambda_m$ from the gate, the working principle is of a two-terminal magnon transistor is similar to that of a three-terminal magnon transistor as long as $d<\lambda_m$. However, the two-terminal magnon transistor the coinciding injector and gate contact give rise to an additional $I_{AC}I_{DC}\cos{\alpha}$ cross term for $V^{1\omega}$ compared to equation (\ref{eq:V1angular}), with $D$ being a prefactor (see SI):
\begin{equation}
    V^{1\omega}=A^{1\omega}\cos^2{(\alpha)} + B^{1\omega}\cos^3{(\alpha)} + D\cos{(\alpha)}.
\end{equation}

In Fig. \ref{fig2:AC_DC} the non-local voltages as a function of in-plane angle $\alpha$ for an injector current $I_{AC}=$ 60 $\mu$A and a DC-current of $I_{DC}=$-50 $\mu$A and $I_{DC}=$-100 $\mu$A are shown at a field of $B=$ 7.5 T. We performed our experiments at $T=$ 25 K as the non-local magnon transport shows a maximum at this temperature\cite{de_wal_long-distance_2023}. In \ref{fig2:AC_DC}a,c the symmetrized, around $\alpha=90^{\circ}$ see SI, (antisymmetrized) non-local voltage responses are shown for $V^{1\omega}$ ($V^{2\omega}$). The fits correspond to the $A^{1\omega}$ ($A^{2\omega}$) term in equation (\ref{eq:V1angular}) ((\ref{eq:V2angular})). In Fig. \ref{fig2:AC_DC}b,d the antisymmetrized, (symmetrized) voltage responses for the first (second) harmonic are given. The fits correspond to the ($D\cos{\alpha}$) term. 

In Fig. \ref{fig2:AC_DC}b the $D\cos{\alpha}$ cross term dominates over the $B^{1\omega}\cos^3{\alpha}$ term. Extraction of $B^{1\omega}$ only yielded values for $B^{1\omega}>A^{1\omega}$, which is nonphysical. 
Yet, in Fig. \ref{fig2:AC_DC}d the effect of $I_{DC}$ via the SHE is clearly observed as the $B^{2\omega}\cos^2{\alpha}$ (the latter part of equation (\ref{eq:V2angular})). As $I_{DC}$ affects the magnon density via the SHE, we expect the same modulation via the SHE in the first harmonic signal $B^{1\omega}$ compared to the second harmonic signal $B^{2\omega}$. 

\begin{figure}
\centering
\includegraphics[width=\linewidth]{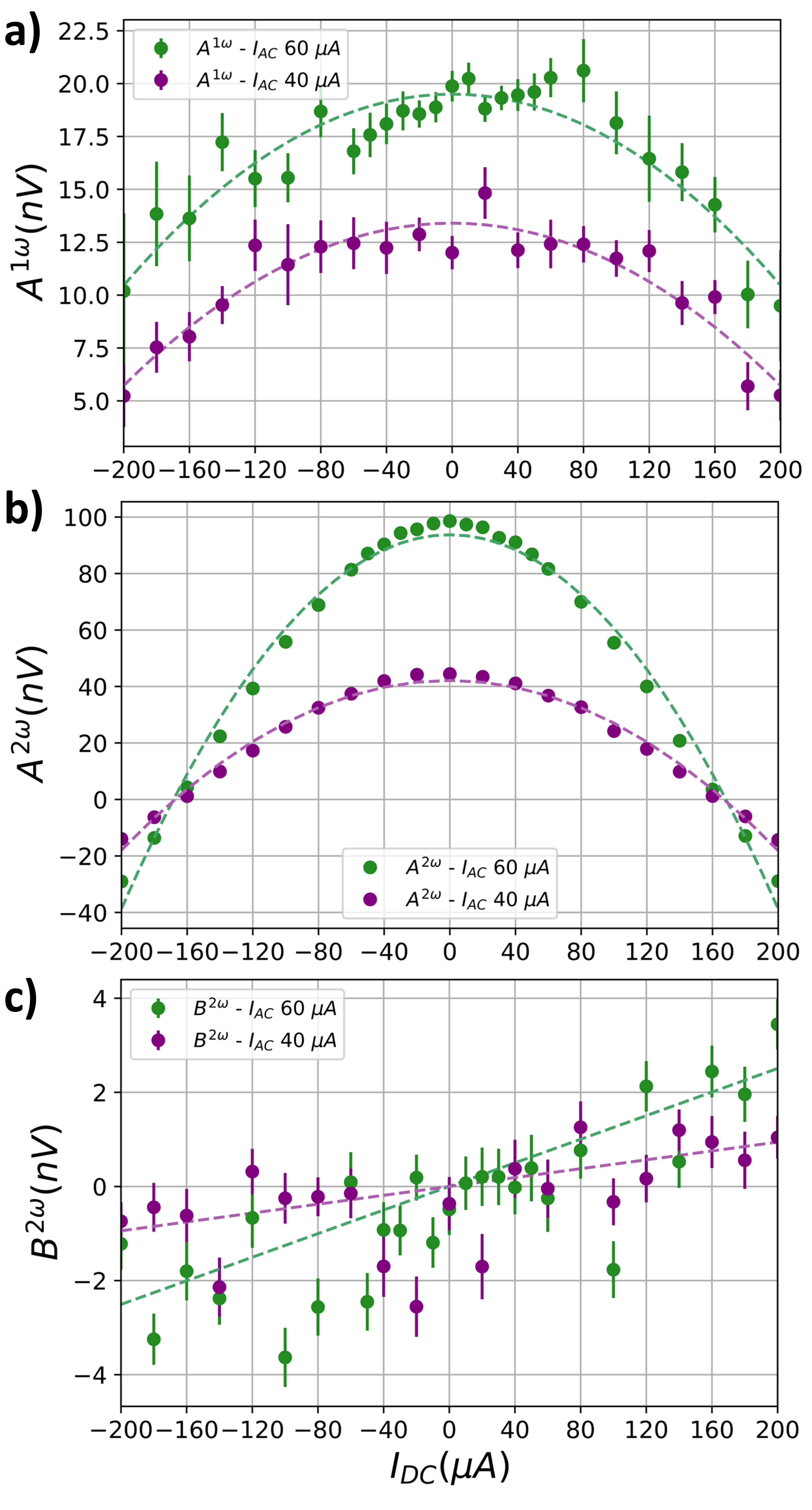}
\caption{Modulation in device D1 of $A^{1\omega}$, $A^{2\omega}$, and $B^{2\omega}$ by $I_{DC}$ via Joule heating (a) and (b) and by the SHE (c). The coefficients given in equation (\ref{eq:V1angular}) and \ref{eq:V2angular} are given as a function of $I_{DC}$. For $I_{AC}=$ 60 $\mu$A (Green) and $I_{AC}=$ 40 $\mu$A (Purple), $A^{1\omega}$ (\textit{top}) and $A^{2\omega}$ (\textit{center}) are fitted, dashed line, and show a quadratic dependence on $I_{DC}$. $B^{2\omega}$ (\textit{bottom}) is fitted and shows a linear dependence on $I_{DC}$. This is expected from eq. \ref{eq:V1angular} and eq. \ref{eq:V2angular}. For $A^{1\omega}$ a linear offset w.r.t. $I_{DC}$ is subtracted (see SI). }
\label{fig3:DCmodulation}%
\end{figure}

Both $A^{1\omega}$ and $A^{2\omega}$ are given as a function of $I_{DC}$ in Fig. \ref{fig3:DCmodulation}a and \ref{fig3:DCmodulation}b for $I_{DC}$ between -200 $\mu$A and +200 $\mu$A. Fit results for $A^{1\omega}$ and $A^{2\omega}$ are indicated by the dashed lines and show a quadratic dependence on $I_{DC}$. The sign of the gate current dependence of $A^{1\omega}$ and $A^{2\omega}$ is equal, indicating that both the electrically and thermally generated magnon spin transport are reduced by an increased magnon temperature. In contrast to the magnon gate on YIG, in which the thermally generated magnon transport is enhanced by the enhanced temperature due to Joule heating by the modulator\cite{cornelissen_spin-current-controlled_2018}, we only see a decrease in $V^{2\omega}_{nl}$. Fig. \ref{fig3:DCmodulation}c, the fit results for $B^{2\omega}$, shows a linear dependence on $I_{DC}$ as expected. The slope $dB^{2\omega}/dI_{DC}$ expresses the modulation efficiency by the SHE injection by $I_{DC}$. At 25 K for a magnetic field of 7.5 T, we find $dB^{2\omega}/dI_{DC}=13\pm2$ nV/mA for $I_{AC}=60$ $\mu$A and $dB^{2\omega}/dI_{DC}=5\pm2$ nV/mA for $I_{AC}=40$ $\mu$A. These values are comparable to the values measured in the three-terminal magnon transistor on YIG\cite{cornelissen_spin-current-controlled_2018}. Comparing the modulation to the zero-gating ($I_{DC}=0$) signals, where $B^{1\omega(2\omega)} = 0$, we can extract the relative efficiency of modulation:
\begin{equation}
    \eta_{\scriptscriptstyle{SHE}} = \frac{dB^{1\omega(2\omega)}/dI_{DC}}{A^{1\omega(2\omega)}_0},
    \label{eq:SHEefficiency}
\end{equation}
where $A^{1\omega(2\omega)}_0=A^{1\omega(2\omega)} (I_{DC}=0)$. We find $\eta_{\scriptscriptstyle{SHE}}=13$ $\pm2\%/$mA for the second harmonic for $I_{AC}=60$ $\mu$A and $\eta_{\scriptscriptstyle{SHE}}=5\pm2$ $\%/$mA for $I_{AC}=40$ $\mu$A. 
The sign of $dB^{2\omega}/dI_{DC}$ is consistent, as a positive (negative) DC-current via the SHE corresponds to an increase (decrease) in $\mu_m$ and therefore in $n_m$ (see Fig. \ref{fig2:AC_DC}d and Fig. \ref{fig3:DCmodulation}c).

In short, the use of the conventional three terminal magnon transistor geometry for CrPS\textsubscript{4} is restricted by the relatively short $\lambda_m$ together with the middle gate contact functioning as a spin sink. However, by exploring a two terminal transistor, where the injector and gate coincide, we show clear modulation by both changing $n_m$ by the SHE as well as by Joule heating. Yet, for the first we can only observe it for the thermally generated non-local magnon transport signal ($V^{2\omega}_{nl}$). 

\begin{figure*}
    \centering
    \includegraphics[width=0.9\linewidth]{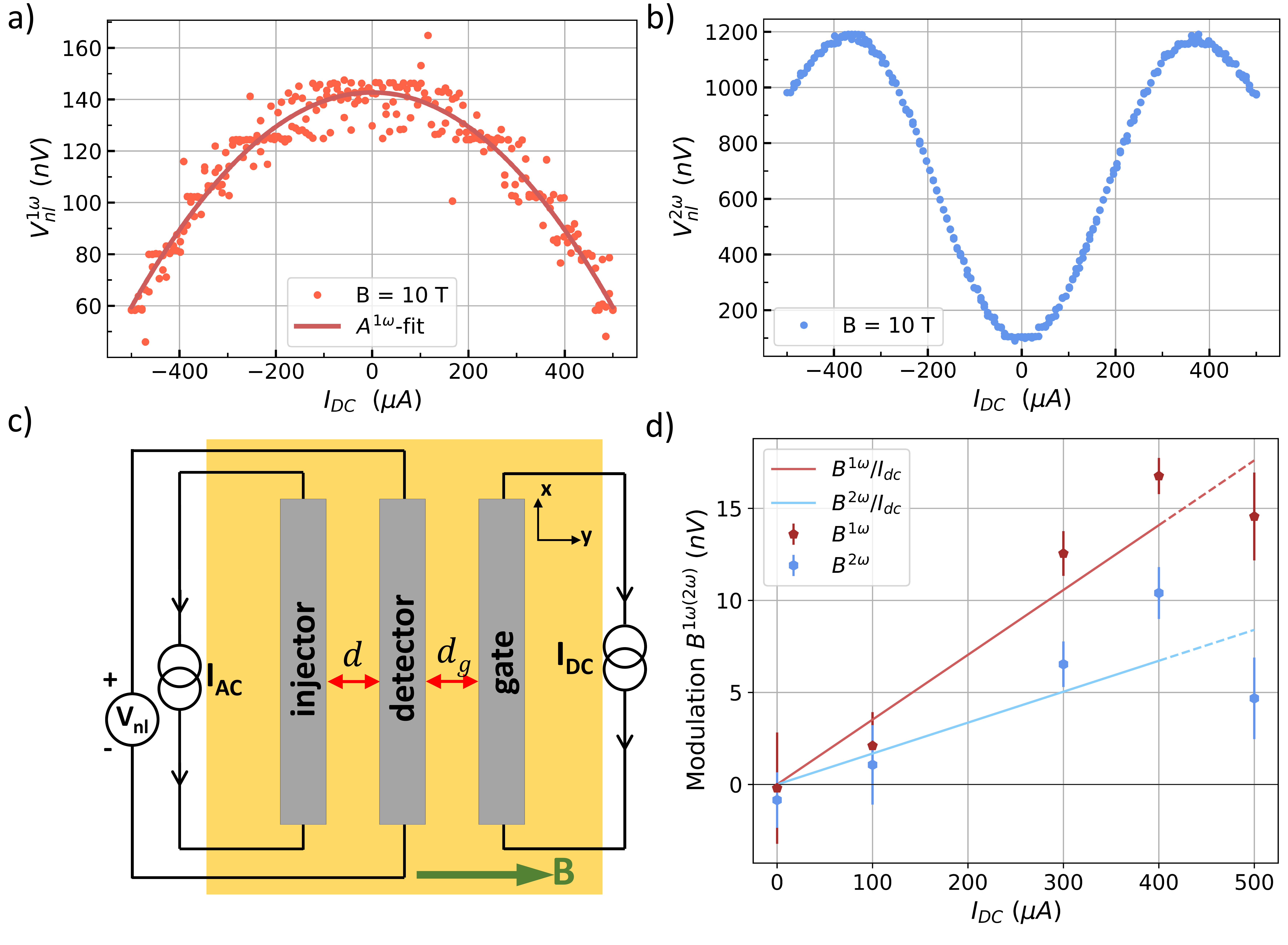}
    \caption{Result from device D2: $I_{DC}$ dependence of a) $V^{1\omega}$ and b) $V^{2\omega}$ at a magnetic field of 10 T at 25 K and $I_{AC}=$ 100 $\mu$A. c) Unconventional three terminal magnon transistor where the modulating gate Pt contact is located on the opposite side of the detector contact compared to the injector contact. The circuitry indicated the connection of the current sources and voltage probes. d) Modulation of $V^{1\omega}$ ($V^{2\omega}$) by the SHE from the gate contact. The modulation amplitude is given in terms of $B^{1\omega(2\omega)}$ as given in equation (\ref{eq:V1modulated}) ((\ref{eq:V2modulated})). This data is obtained from antisymmetrizing (symmetrizing) w.r.t. field, see SI.}
    \label{fig:4}
\end{figure*}

In addition to this we measured a second device D2. As a proof of principle we investigated another unconventional magnon transistor geometry, where the injector-detector contacts are directly adjacent and the gate contact is located on the opposite side of the detector compared to the injector. This is illustrated in Fig. \ref{fig:4}c. The Pt strips are aligned with the \textit{a}-crystal-axis of the CrPS\textsubscript{4}, such that the magnon transport channel is along the b-axis. Both the electrically and thermally generated magnon transport channels can be modulated by the gate contact as the injected magnon spin and heat can diffuse towards the magnon transport channels, even though the gate contact is outside these channels, as long as $d_g\ll\lambda_m$ (where $d_g$ is the distance between the detector and gate). In Fig. \ref{fig:4}a and \ref{fig:4}b the first and second harmonic non-local voltages at an in plane field of 10 T in the y-direction at 25 K, with $I_{AC}=$ 100 $\mu$A for device D2, are shown as a function of $I_{DC}$ with the modulator strip as depicted in Fig. \ref{fig:4}c. The edge-to-edge distance $d$ between the injector and detector contact is 270 nm and the distance $d_g$ between the detector and gate contact is 480 nm. Both $d\ll \lambda_m$ and $d_g\ll \lambda_m$, where $\lambda_m$ is the magnon relaxation length of CrPS\textsubscript{4}. At 10 T the sub-lattice magnetizations of CrPS\textsubscript{4} are saturated in plane and aligned with the magnetic field (see \cite{de_wal_long-distance_2023}). The first harmonic voltage response is fitted (solid red line in Fig. \ref{fig:4}a) and shows a quadratic dependence of $A^{1\omega}$ on $I_{DC}$, as is the case for device D1 (Fig. \ref{fig3:DCmodulation}a). 

In contrast, the second harmonic voltage response, given in Fig. \ref{fig:4}b, shows a very different modulation as a function of $I_{DC}$. To our surprise, $V^{2\omega}_{nl}$ is strongly enhanced by $I_{DC}$ up to one order of magnitude at $I_{DC}\approx\lvert370$ $\mu$A$\rvert$ and at larger $I_{DC}$, $V^{2\omega}_{nl}$ decreases. This is different than has been observed in conventional three terminal magnon valve systems on MnPS\textsubscript{3}\cite{chen_electrically_2021}, CrPS\textsubscript{4}\cite{qi_giant_2023}, and even in the two-terminal geometry in device D1 (Fig. \ref{fig3:DCmodulation}b), where only a strong suppression of the $V^{2\omega}_{nl}$ is observed and at sufficiently large $I_{DC}$ the zero crossing is realized. Although the temperature dependence of the nl-SSE in CrPS\textsubscript{4} is not fully understood, such a strong enhancement of nl-SSE voltage is at least striking and possibly opens up a completely new route towards the control of magnon spin transport. The absence of this enhanced transport in $V^{1\omega}_{nl}$ shows that the enhancement in $V^{2\omega}_{nl}$ is likely a combined effect of changing $n_m$ via the temperature and the SSE both originating from the Joule heating by $I_{DC}$. Possibly, the thermal gradient originating from the gate contacts enhances the nl-SSE close to the detector contact by altering the thermal gradients generated by the Joule heating in the injector. The Joule heating from the gate current also raises the temperature at the detector strip. The latter suppresses the magnetization at larger gating currents.

Lastly, we compared the effect of $I_{DC}$ via the SHE altering the $B$ parameters in equations (\ref{eq:V1angular}) and (\ref{eq:V2angular}). In Fig. \ref{fig:4}d the dependence of $B^{1\omega(2\omega)}$ for the first harmonic (second harmonic) at an external field of 12 T is plotted as function of $I_{DC}$ (extracted from field dependent measurements, see SI). The fit shows a linear dependence on $I_{DC}$ which is consistent with the results on device D1. Albeit, the modulation efficiency $dB/dI_{DC}$ achieved is larger. Moreover, we found $dB/dI_{DC}$ depends on the magnitude of the magnetic field applied, which is likely caused by the not fully saturated $V^{1\omega}_{nl}$ for larger $I_{DC}$, due to the Joule heating, and the not understood behavior of the nl-SSE as a function of temperatures (see SI). For the efficiency of modulation at 12 T in device D2, we found $\eta^{1\omega}_{\scriptscriptstyle{SHE}}=25$ $\%/$mA ($\eta_{\scriptscriptstyle{SHE}}^{2\omega}=16$ $\%/$mA) for the first (second) harmonic. These values are slightly larger than found for device D1 which is likely caused by the stronger magnetic field. Compared to the modulation efficiencies found in YIG, the values here are 3-5 times larger for device D1 and 5-8 times larger for device D2\cite{cornelissen_spin-current-controlled_2018}. Future study of these unconventional geometries on YIG can further reveal their potential for magnon transistors. 

\section{Conclusion}
Summarizing, we demonstrate control over the magnon spin transport in two magnon transistors with unconventional non-local geometries. 
In these transistors we modulated the electrically and thermally generated magnon spin currents, by using a Pt gate contact which injects magnon spins via the SHE, and by altering the magnon temperature and via the SSE, both by Joule heating. Moreover, we separate the effects of modulation on both the electrically and thermally generated magnon spin currents. In device D1, a two terminal transistor where the injector and gate contact coincide, we find a modulation efficiency by the SHE of $\eta_{\scriptscriptstyle{SHE}}^{2\omega}=13\pm2$ $\%/$mA for the thermally generated magnon current at 7.5 T, 25 K. In device D2, a three terminal magnon transistor with the gate electrode on the opposite side of the detector compared to the injector, we find modulation efficiencies by the SHE of $\eta^{1\omega}_{\scriptscriptstyle{SHE}}=25$ $\%/$mA and $\eta^{2\omega}_{\scriptscriptstyle{SHE}}=16$ $\%/$mA at 12 T, 25 K, for the electrically and thermally generated magnon currents, respectively. These values are 3-8 times larger than modulation efficiencies found in YIG\cite{cornelissen_spin-current-controlled_2018}. Moreover, we show that via altering the temperature of device the thermally generated magnon current can be enhanced by one order of magnitude. These results, pave the way for valorization of magnon spin transport in antiferromagnets in technological applications and contribute to our understanding of controlling magnon spin transport.

\begin{acknowledgments}
We acknowledge the technical support from J. G. Holstein, H. Adema, H. H. de Vries, and F. H. van der Velde. We acknowledge the financial support of the Zernike Institute for Advanced Materials and the European Union’s Horizon 2020 research and innovation program under Grant Agreements No. 785219 and No. 881603 (Graphene Flagship Core 2 and Core 3). This project is also financed by the NWO Spinoza prize awarded to B.J.W. by the NWO and has received funding from the European Research Council (ERC) under the European Union’s 2DMAGSPIN (Grant Agreement No. 101053054). 
\end{acknowledgments}

\bibliography{Reference}

\pagebreak
\widetext








\setcounter{figure}{0}
\renewcommand{\thefigure}{S\arabic{figure}}
\renewcommand{\theequation}{S\arabic{equation}}
\setcounter{section}{0}
\renewcommand{\thesection}{S\arabic{section}}

\begin{center}
\Huge
   \textbf{\textbf{Supplementary Information}}\\
   \Large {\setstretch{1.0} Gate control of magnon spin transport in unconventional magnon transistors based on the van der Waals antiferromagnet CrPS\textsubscript{4} }
\end{center}

\section{The effect of a DC current on the voltage response}
For measuring the voltage response of our system we employ lock-in technique. This allows for the detection of small signals down to nanovolts on noise backgrounds. Furthermore, the simultaneous detection of higher order harmonics is possible. 

For a driving ac-current $I(t)=\sqrt{2}I_0\sin{(\omega t)}$, where $I_0$ is the RMS value of the ac-current and $\omega=2\pi f$ with $f$ being the ac frequency, the voltage response can be extended as follows:
\begin{equation}
    V(t) = R_1I(t) + R_2I^2(t) + ...
    \label{eq:LockinVoltage}
\end{equation}
where only the first and second-order term are shown and $R_1$ and $R_2$ are the coefficients for these terms, respectively. Neglecting higher order terms, the voltages detected for the first harmonic ($V^{1\omega}$) and the second harmonics ($V^{2\omega}$) are then given by:
\begin{align} 
V^{1\omega} =& I_0R_1                       &     &(\phi=0^{\circ}), \label{eq:LockinV1}\\ 
V^{2\omega} =& \frac{I^2_0}{\sqrt{2}}R_2    &     &(\phi=-90^{\circ}). \label{eq:LockinV2}
\end{align}
Here, $\phi$ is the phase of the detected w.r.t. the reference signal. 

In our experiment we apply a dc-current on top of the driving ac-current. The effect of this dc-current does not simply give an offset to the detected voltage response. The initial current can be extended with a dc term: 
\begin{equation}
    I(t)=\sqrt{2}I_0\sin{(\omega t)} + I_{DC}.
    \label{eq:DConACcurrent}
\end{equation}
In order to investigate the possible effect on any cross-terms we substitute equation \ref{eq:DConACcurrent} into equation \ref{eq:LockinVoltage} to find the voltage responses:
\begin{align*}
    V(t)&=R_1\left[\sqrt{2}I_0\sin{(\omega t)}+I_{DC}\right] + R_2\left[(\sqrt{2}I_0\sin{(\omega t)}+I_{DC})^2\right]\\
    &=R_1\left[\sqrt{2}I_0\sin{(\omega t)}+I_{DC}\right] + R_2\left[(\sqrt2{I_0})^2\sin^2{(\omega t)}+2\sqrt{2}I_0I_{DC}\sin{(\omega t)}+I^2_{DC}\right]
\end{align*}
\newline
$V^{1\omega}$ and $V^{2\omega}$ are only sensitive to the ac-contributions, hence the $I_{DC}$ and $I^2_{DC}$ terms are ignored:
\begin{align*}
    V(t) &=R_1\left[\sqrt{2}I_0\sin{(\omega t)}\right] + R_2\left[2\sqrt{2}I_0I_{DC}\sin{(\omega t)}\right]
    + R_2\left[2I^2_0\left(\frac{1-\cos{(2\omega t)}}{2}\right)\right]\\
    &=\sqrt{2}I_0\sin{(\omega t)}\big[R_1 + 2R_2I_{DC}\big] + 
    R_2\left[I^2_0+I^2_0\sin{\left(2\omega t-\frac{\pi}{2}\right)}\right]
\end{align*}
\newline
In this, the $R_2I_{0}^2$ term can be ignored as well such that we obtain (Note that the $sin(2\omega t - \frac{\pi}{2})$ term is the -90$^{\circ}$ shift in eq. \ref{eq:LockinV2} and eq. \ref{eq:LockinV2DC}):
\begin{align} 
V^{1\omega} =& I_0\cdot(R_1 + 2R_2I_{DC}) & &(\phi=0^{\circ}), \label{eq:LockinV1DC}\\ 
V^{2\omega} =& \frac{I^2_0}{\sqrt{2}}R_2 & &(\phi=-90^{\circ}). \label{eq:LockinV2DC}
\end{align}
Compared to the voltage response without a DC bias (equations \ref{eq:LockinV1} and \ref{eq:LockinV2}), only the first harmonic response, $V^{1\omega}$ (equation \ref{eq:LockinV1DC}), has been altered. 

The effect of the additional term in equation \ref{eq:LockinV1DC} can have a major impact on the first harmonic voltage response, in the case $R_1\ll 2R_2I_{DC}$. As this is the case for our system, the large contribution from the cross-term, $2R_2I_{DC}$, overshadows the electrically injected (SHE) magnon signal and the electrical modulation via the SHE, especially at large gate currents $I_{DC}$.

Therefore, equation 7 in the main text, which is altered by this cross-term as follows,
\begin{equation}
    V^{1\omega} = A^{1\omega}\cos^2{(\alpha)} + B^{1\omega}\cos^3{(\alpha)} + D \cos{(\alpha)},
    \label{eq:V1total}
\end{equation}
where the last term is added by the angular dependence $2R_2I_{DC}$, thus showing the same angular dependence as the second harmonic voltage response. 

Separating the $D \cos{(\alpha)}$ component from the $B^{1\omega}\cos^3{(\alpha)}$ component yielded a value for $B^{1\omega}\gg A^{1\omega}$, which is nonphysical. We cannot fully explain this behavior. Possibly the in plane anisotropy of CrPS\textsubscript{4} effects the magnon transport and therefore the non-local voltages.

\subsection{Additional terms in $V^{1\omega}$ due to the injector and gate coinciding}
With similar reasoning for arriving at the $D\cos{(\alpha)}$ term, the $2R_2I_{DC}$ term in equation \ref{eq:LockinV1DC} will also lead to another additional term: $E\cos^2{(\alpha)}$ in equation \ref{eq:V1extra}, which arise due to $B^{2\omega}\cos^2{(\alpha)}$:
\begin{align}
        V^{1\omega} &= A^{1\omega}\cos^2{(\alpha)} + E\cos^2{(\alpha)} + B^{1\omega}\cos^3{(\alpha)} + D \cos{(\alpha)}\label{eq:V1extra}\\
            &= (A^{1\omega} + E)\cos^2{(\alpha)} + B^{1\omega}\cos^3{(\alpha)} + D \cos{(\alpha)}
\end{align}
The $E\cos^2{(\alpha)}$ term, which depends linearly on $I_{DC}$, like $B^{2\omega}$, will simply give an offset ($E$) for $A^{1\omega}$ extracted in figure 2 and 3 in the main text, where $E\propto I_{DC}$. In figure 3a in the main text the plotted $A^{1\omega}$ is corrected for this cross term.


\section{Symmetry analysis}
The different components of the first and second harmonic voltage responses in figure 2 in the main text, as given in equation 5-7 in the main text, are extracted by symmetrizing the measured $V^{1\omega}_{nl}$ and $V^{2\omega}_{nl}$ around $\alpha=90^{\circ}$. All ADMR measurements are performed in trace ($\alpha$ from 0 to 360$^{\circ}$) and retrace ($\alpha$ from 360$^{\circ}$ to 0). Symmetrizing is achieved by shifting the retrace data by 180$^{\circ}$ and adding it to the trace data, after which the data is normalized. Antisymmetrizing is achieved by subtraction instead of addition of the shifted retrace data.

\begin{figure}[t!]
    \centering
    \includegraphics[width=\linewidth]{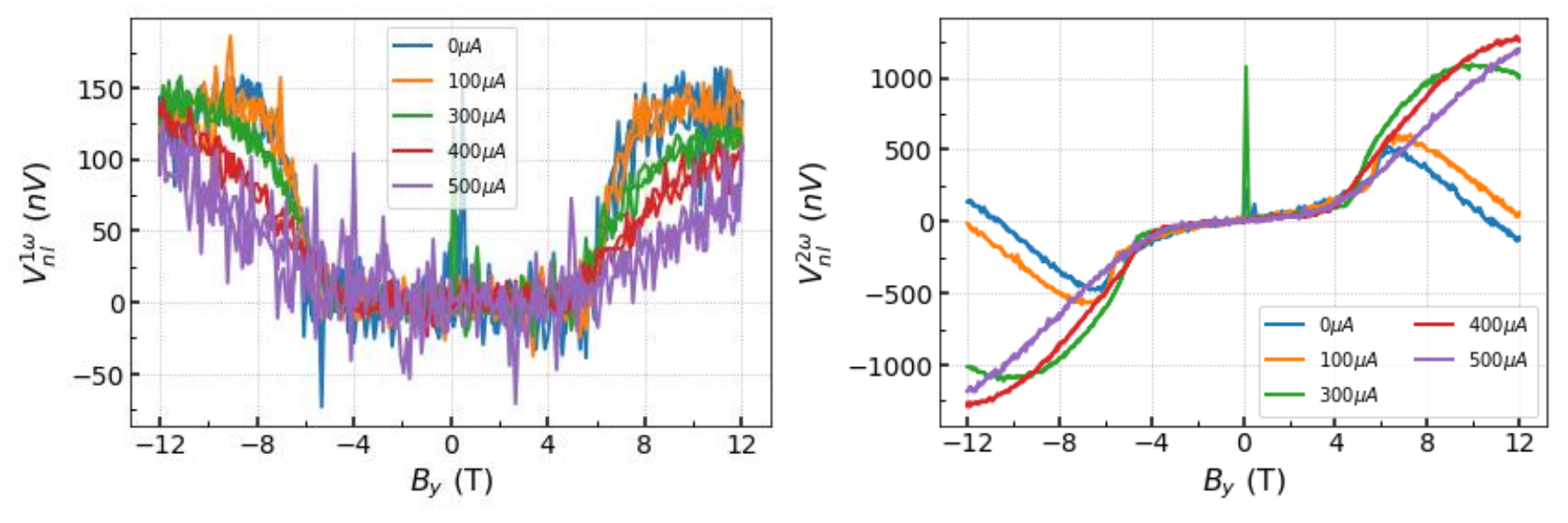}
    \caption{Magnetic field dependence of $V^{1\omega}_{nl}$ and $V^{2\omega}_{nl}$ for device D2. a) $V^{1\omega}_{nl}$ data at different $I_{DC}$ (as indicated in the legends) for at 25 K for $I_{ac}=100\mu$A. An offset value of 74 nV has been subtracted. b) $V^{2\omega}_{nl}$ data measured simultaneously with $V^{1\omega}_{nl}$, no offset has been removed. The peak at 0 T we consider an outlier.}
    \label{fig:S3.1}
\end{figure}

\section{Magnetic field dependence of $V^{1\omega}$ and $V^{2\omega}$ for the unconventional three-terminal magnon transistor (device D2)}
\begin{figure}[h]
    \centering
    \includegraphics[width=\linewidth]{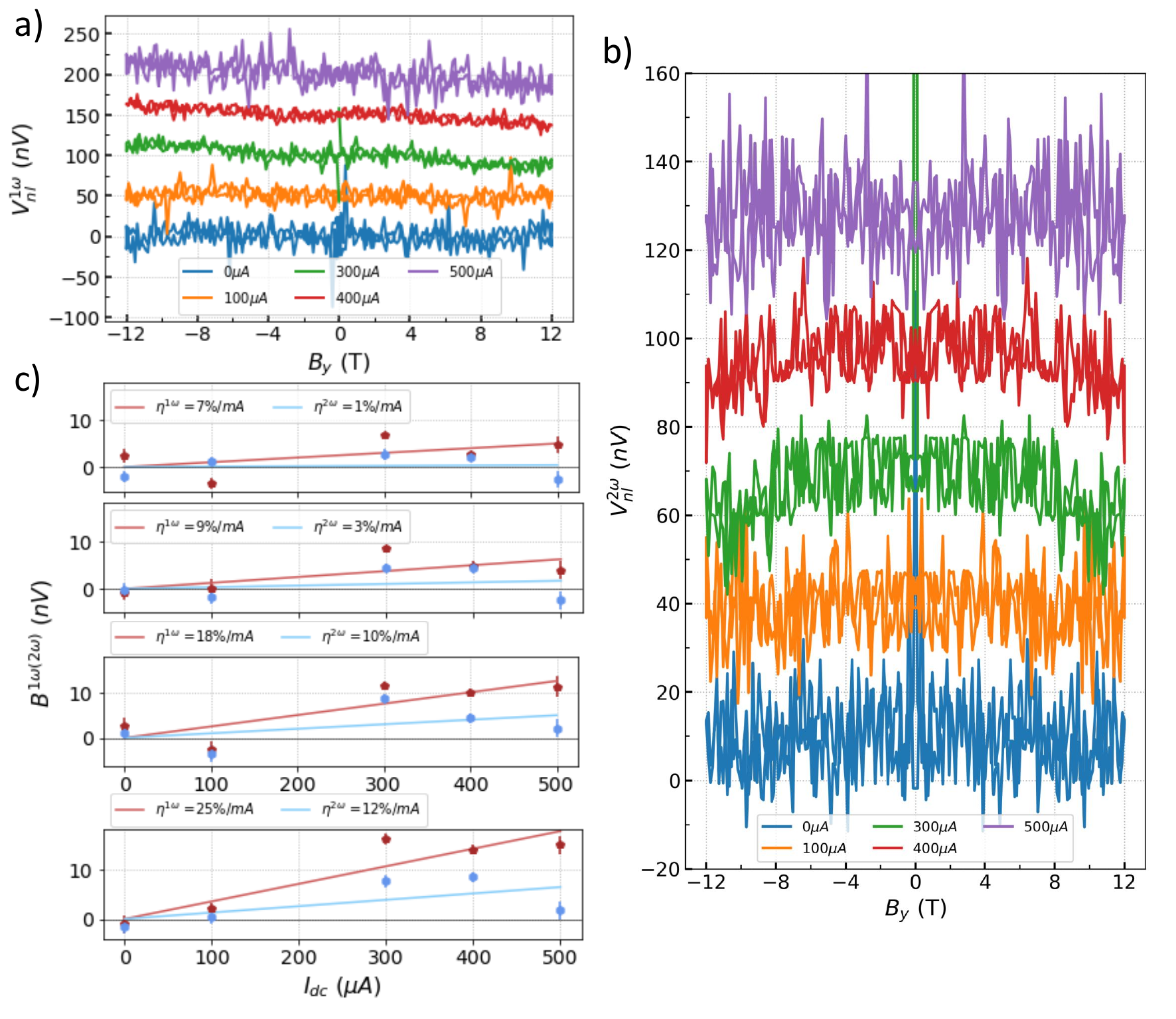}
    \caption{a) Antisymmetrized $V^{1\omega}$ signal as function of field, to extract $B^{1\omega}$ and b) Symmetrized $V^{2\omega}$ to extract $B^{2\omega}$ at different $I_{DC}$. The data for the different $I_{DC}$ is spaced by 50 nV for $V^{1\omega}$ and by 30 nV for $V^{2\omega}$. c) $B^{1\omega(2\omega)}$ extracted from a) (and b)) at different magnetic field strenghts: From top to bottom; 8 T, 9 T, 10 T, 11 T. The modulation efficiency $\eta^{1\omega(2\omega)}=(dB^{1\omega(2\omega)}/I_{DC} )/ A^{1\omega(2\omega)}_0$ is given in the legends. }
    \label{fig:S3.2}
\end{figure}

The modulation efficiencies ($\eta^{1\omega}$ and $\eta^{2\omega}$) in the unconventional three terminal magnon transistor device D2 (given in figure 4c in the main text) are extracted via symmetry analysis. The raw data for both $V^{1\omega}_{nl}$ and $V^{2\omega}_{nl}$ are given in figure \ref{fig:S3.1}. In figure \ref{fig:S3.2}a the antisymmetric part (w.r.t. field) of $V^{1\omega}_{nl}$ is depicted. Here symmetrizing/antisymmetrizing is achieved by addition/subtraction from the data measured at positive fields and negative fields. For $V^{1\omega}$ given in equation 2 in the main text, the measured $V^{1\omega}_{nl}$, excited by $I_{ac}$, is symmetric in field. The modulation by $I_{DC}$ via the SHE ($B^{1\omega}\cos^3{(\alpha)}$) is asymmetric in field (see equation 5 in the main text). Therefore, this modulation can be extracted by antisymmetrizing $V^{1\omega}$, which is shown in \ref{fig:S3.2}a. Equal reasoning can be applied to $V^{2\omega}$, however, for $V^{2\omega}$ the symmetric part corresponds to $B^{2\omega}\cos^2{(\alpha)}$ (see equation 3 and 6) in the main text). Therefore, we extract $B^{2\omega}$ from the symmetrized $V^{2\omega}$ as is shown in figure \ref{fig:S3.2}b. 

In figure \ref{fig:S3.2}c $B^{1\omega(2\omega)}$ extracted from the antisymmetrized (symmetrized) $V^{1\omega}_{nl}$ and $V^{2\omega}_{nl}$ are given for different magnetic field strenghts. The $B^{1\omega(2\omega)}$ at specific field strenghts $B_y$ are obtained by taking the average of $V^{1\omega(2\omega)}_{nl}$ in the window [$B_y-0.5T$, $B_y+0.5T$], the standard error is indicated by the error bars. The slope of the linear fit, which is $B^{1\omega(2\omega)}/I_{DC}$ is used to calculated the modulation efficiencies, as indicated in the legends of figure \ref{fig:S3.2}c.

\newpage


\end{document}